\documentclass[12pt]{article}
\usepackage{mathrsfs}
\usepackage{graphicx}
\usepackage{amsmath}
\usepackage{amssymb}
\usepackage{caption2}
\begin{document}
\newcommand  {\ba} {\begin{eqnarray}}
\newcommand  {\be} {\begin{equation}}
\newcommand  {\ea} {\end{eqnarray}}
\newcommand  {\ee} {\end{equation}}
\renewcommand{\thefootnote}{\fnsymbol{footnote}}
\renewcommand{\figurename}{Figure.}
\renewcommand{\captionlabeldelim}{.~}

\vspace*{1cm}
\begin{center}
 {\Large\textbf{Fermion Masses and Leptogenesis from The Left-Right Symmetric Model}}

\vspace{1cm}
 \textbf{Wei-Min Yang}

\vspace{0.4cm}
 \emph{Department of Modern Physics, University of Science and Technology of China, Hefei 230026, P. R. China}

\vspace{0.2cm}
 \emph{E-mail: wmyang@ustc.edu.cn}
\end{center}

\vspace{1cm}
\noindent\textbf{Abstract}: The paper suggests a left-right mirror symmetric model with the flavor symmetries $Z_{3L}\otimes Z_{3R}$. It can simultaneously accommodate the standard model, neutrino physics and the baryon asymmetry. The fermion masses and $CP$ violation originate from vacuum spontaneous breaking of the flavor field. The baryon asymmetry is implemented through the leptogenesis which is related to the lepton $CP$ violation. The model can naturally and correctly reproduce all kinds of experimental data, in particular, all of the values of the fermion masses and mixings are accurately fitted by the fewer parameters. Finally, it is also feasible and promising to test the model in future experiments.

\vspace{1cm}

\vspace{0.3cm}

\newpage
 \noindent\textbf{I. Introduction}

\vspace{0.3cm}
 The standard model (SM) has been evidenced to be a correct theory at the electroweak scale, but it also contains some flaws \cite{1}. The left-handed and right-handed fermions belong to the asymmetric gauge group representations. The Yukawa couplings have too large freedoms in view of the lack of a flavor symmetry, as a result, the fermion masses and mixings are apparently in confusion, and the origin of the $CP$ violations is still unknown, this is namely so-called flavor puzzle \cite{2}. On the other hand, the SM can not account for such issues as the light neutrino masses \cite{3}, the baryon asymmetry \cite{4}, the cold dark matter \cite{5}, etc. Therefore, the SM should only be regarded as a low-energy effective theory, there must be a more full and fundamental theory beyond the SM, which is ultimately in charge of the matter origin in the early universe evolution.

 In the last few decades, all the time all kinds of theoretical ideas have been suggested to solve the above-mentioned issues. The left-right mirror symmetry is a well-motivated idea because it meets the aesthetics. The left-right symmetric model is pioneered by Pati and Salam in \cite{6}, later is developed by the authors in \cite{7}. This type of model can be derived from the GUT and have many advantages, so they have been extensively studied \cite{8}. Recently, the researches of the flavor symmetry attract great attentions in view of it succeeding in the neutrino mass and mixing \cite{9}. The flavor symmetry is surely relevant to the origin of the fermion masses and $CP$ violation, this connection has been discussed in a lot of references \cite{10}. The baryon asymmetry through the leptogenesis is also a successful idea \cite{11}, many progresses have been made in this field \cite{12}. These ideas are all insights and can be considered as approaches to a new theory. However, the new theory should in line with the early universe harmony and the nature unification. Therefore, a realistic theory should simultaneously accommodate the SM, neutrino physics and the baryon asymmetry, moreover, it can account for the origin of the fermion masses and $CP$ violation, in other words, the flavor symmetry has to be integrated into it. Undoubtedly, exploring such theoretical model is very significant for particle physics as well as cosmology.

 In this work, I suggest a left-right symmetric model. The model has the local gauge groups $SU(3)_{C}\otimes SU(2)_{L}\otimes SU(2)_{R}\otimes U(1)_{B-L}$ and the flavor symmetries $Z_{3L}\otimes Z_{3R}$. Besides the SM particles, some new particles are introduced in the model. The model symmetries and their breakings lead to a special effective Yukawa couplings at the low energy, essentially, the fermion masses and mixings and the $CP$ violations originate from vacuum spontaneous breaking of the flavor field. The generated quark and lepton mass matrices are not independent but rather interrelated. In addition, the baryon asymmetry is successfully implemented through the leptogenesis, which is closely related to the lepton $CP$ violation. The model can correctly reproduce all kinds of the observed data of the SM, neutrino physics and the baryon asymmetry, in particular, all of the values of the fermion masses and mixings are accurately fitted by the fewer parameters. Finally, the model is feasible and promising to be tested in future experiments.

 The remainder of this paper is organized as follows. In Section II I outline the model and discuss the fermion mass generations. Sec. III I introduce the leptogenesis in the model. The numerical results are given in Sec. IV. Sec. V is devoted to conclusions.

\vspace{1cm}
 \noindent\textbf{II. Model}

\vspace{0.3cm}
 The gauge symmetries of the model are the left-right symmetric local groups $SU(3)_{C}\otimes SU(2)_{L}\otimes SU(2)_{R}\otimes U(1)_{B-L}$. The model particle contents and their gauge quantum numbers are in detail listed as follows,
\begin{alignat}{1}
 &G_{\mu}(8,1,1,0),\hspace{0.5cm} W_{L\mu}(1,3,1,0),\hspace{0.5cm} W_{R\mu}(1,1,3,0),\hspace{0.5cm}
  X_{\mu}(1,1,1,1),\nonumber\\
 &[\Phi_{L},q_{L}](3,2,1,\frac{1}{3}),\hspace{0.3cm} [\Phi_{R},{q}_{R}](3,1,2,\frac{1}{3}),\hspace{0.3cm} [H_{L},l_{L}](1,2,1,-1),\hspace{0.3cm} [H_{R},{l}_{R}](1,1,2,-1),\nonumber\\
 &\phi_{c}(3,1,1,-\frac{2}{3}),\hspace{0.5cm} \phi^{-}(1,1,1,-2),\hspace{0.5cm} [N_{L},N_{R}](1,1,1,0),\hspace{0.5cm} F(1,1,1,0),
\end{alignat}
 where all kinds of the fermions imply three generations as usual. $H_{L}$ and $H_{R}$ are respectively the left-handed and right-handed isospin doublet scalars, while $\Phi_{L}$ and $\Phi_{R}$ are scalars which belong to both color triplet and isospin doublet. $N_{L}$ and $N_{R}$ are singlet Majorana fermions. $F$ is a singlet scalar flavor field, which is a $3\times3$ matrix in the flavor space. The other notations are self-explanatory. In addition to the gauge symmetries, the model has an attractive left-right mirror symmetry. It is characterized by the field transforms as follows,
\ba
 W_{L\mu}\leftrightarrow W_{R\mu},\hspace{0.5cm} \Phi_{L}/H_{L}\leftrightarrow \Phi_{R}/H_{R},\hspace{0.5cm} q_{L}/l_{L}/N_{L}\leftrightarrow q_{R}/l_{R}/N_{R},\hspace{0.5cm} F\leftrightarrow F^{\dagger},
\ea
 where the mirror particles of $G_{\mu},X_{\mu},\phi_{c},\phi^{-}$ are exactly themselves. Lastly, the model has the flavor symmetries $Z_{3L}\otimes Z_{3R}$, which are three order cyclic groups for three generations of the left-handed and right-handed fermions, respectively. The above-mentioned symmetries are simple and natural, they are the theoretical basis of the model.

 The model Lagrangian is easy written out on the basis of the symmetries and particle contents. The gauge kinetic energy terms are
\begin{alignat}{1}
 \mathscr{L}_{Gauge}=
 &\:\mathscr{L}_{pure\,gauge}+i\overline{f_{L}}\gamma^{\mu}D_{\mu}f_{L}
  +i\overline{f_{R}}\gamma^{\mu}D_{\mu}f_{R}\nonumber\\
 &+(D^{\mu}\Phi_{L})^{\dagger}D_{\mu}\Phi_{L}+(D^{\mu}\Phi_{R})^{\dagger}D_{\mu}\Phi_{R}
  +(D^{\mu}H_{L})^{\dagger}D_{\mu}H_{L}+(D^{\mu}H_{R})^{\dagger}D_{\mu}H_{R}\nonumber\\
 &+(D^{\mu}\phi_{c})^{\dagger}D_{\mu}\phi_{c}+(D^{\mu}\phi^{-})^{\dagger}D_{\mu}\phi^{-}
  +Tr[\partial^{\mu}F^{\dagger}\partial_{\mu}F],
\end{alignat}
 where $f_{L},f_{R}$ denote the left-handed and right-handed fermions, and the covariant derivative $D_{\mu}$ is defined by
\ba
 D_{\mu}=\partial_{\mu}+i\left(g_{s}G_{\mu}^{a}\frac{\lambda^{a}}{2}+g_{w}W_{L\mu}^{i}\frac{\tau_{L}^{i}}{2}+g_{w}W_{R\mu}^{i}\frac{\tau_{R}^{i}}{2}+g_{x}X_{\mu}\frac{B-L}{2}\right).
\ea
 $g_{s},g_{w},g_{x}$ are three gauge coupling parameters. $\lambda^{a}$ and $\tau^{i}$ are respectively the Gell-Mann and Pauli matrices. $B-L$ is the charge operator of $U(1)_{B-L}$, namely the baryon number minus the lepton one.

 The model Yukawa couplings are
\begin{alignat}{1}
 \mathscr{L}_{Yukawa}=
 &\:H_{L}^{\dagger}N_{L}^{T}CY_{1}l_{L}+\Phi_{L}^{\dagger}N_{L}^{T}CY_{2}q_{L}+\phi_{c}^{\dagger}l_{L}^{T}CY_{3}i\tau_{2}q_{L}
  +\frac{1}{2}\phi^{+}l_{L}^{T}CY_{4}i\tau_{2}l_{L}\nonumber\\
 &+H_{R}^{\dagger}N_{R}^{T}CY_{1}l_{R}+\Phi_{R}^{\dagger}N_{R}^{T}CY_{2}q_{R}
  +\phi_{c}^{\dagger}l_{R}^{T}CY_{3}i\tau_{2}q_{R}+\frac{1}{2}\phi^{+}l_{R}^{T}CY_{4}i\tau_{2}l_{R}\nonumber\\
 &-\frac{1}{2}N_{L}^{T}CM_{N}N_{L}-\frac{1}{2}N_{R}^{T}CM_{N}N_{R}-\overline{N_{L}}Y_{5}^{\dagger}FY_{5}N_{R}+h.c.\,,
\end{alignat}
 where $C$ is a charge conjugation matrix, $i\tau_{2}$ is inserted in order to satisfy the $SU(2)$ isospin symmetries. The left-right mirror symmetry is very evident. The model flavor symmetries are characterized by $Z_{3L}\otimes Z_{3R}$ as follows,
\begin{alignat}{1}
 &T=\left(\begin{array}{ccc}0&0&1\\1&0&0\\0&1&0\end{array}\right),\hspace{0.3cm} T^{2}=T^{-1}=T^{T},\hspace{0.3cm}
  T^{3}=I,\hspace{0.3cm} T^{T}Y_{k}T=Y_{k},\nonumber\\
 &Y_{k}=a_{k}T+b_{k}T^{T}+c_{k}I,\hspace{0.3cm} M_{N}=\overline{M}_{N}(T+T^{T}+c_{N}I)=\overline{M}_{N}Y_{N},\nonumber\\
 &F=U_{FL}\mathrm{diag}(F_{1},F_{2},F_{3})U_{FR}^{\dagger},\nonumber\\
 &f_{L}\longrightarrow T_{L}f_{L},\hspace{0.5cm} f_{R}\longrightarrow T_{R}f_{R},\hspace{0.5cm}
  F\longrightarrow T_{L}F\,T_{R}^{T},
\end{alignat}
 where $k=1,2,3,4,5$. $T$ is the generator of $Z_{3}$ and $I$ is an unit matrix. $Y_{k}$ and $M_{N}$ have simple structures and  fewer parameters by virtue of the flavor symmetries. all of the coefficients, $a_{k},b_{k},c_{k},c_{N}$, are required to be real numbers, so the $CP$ invariance is also satisfied. $Y_{4}$ is an antisymmetric matrix on account of $\tau_{2}^{T}=-\tau_{2}$, so $a_{4}=-b_{4}, c_{4}=0$. $Y_{N}$ is a symmetric matrix. $\overline{M}_{N}$ should be the GUT scale, namely $\overline{M}_{N}\sim 10^{15}$ GeV. The flavor field $F$ is parameterized by three complex scalar fields and two dimensionless unitary flavor fields. The left-handed and right-handed fermions are independently transformed according to the separate group $Z_{3L}$ and $Z_{3R}$, so there are not directly couplings of the left-handed fermions to the right-handed ones except the last term in (5), which is the only link between the left-handed fermions and the right-handed ones. These characteristics of the Yukawa couplings will result in a new mechanism by which the fermion masses and mixings and the $CP$ violation are generated.

 The model scalar potentials are given by
\begin{alignat}{1}
 V_{Scalar}=&\:\lambda_{H}\left(H_{L}^{\dagger}H_{L}-\frac{\lambda_{H}v_{L}^{2}+\lambda_{1}v_{R}^{2}}{2\lambda_{H}}\right)^{2}
  +\lambda_{H}\left(H_{R}^{\dagger}H_{R}-\frac{\lambda_{H}v_{R}^{2}+\lambda_{1}v_{L}^{2}}{2\lambda_{H}}\right)^{2}\nonumber\\
 &+\lambda_{F}\left(Tr[F^{\dagger}F]-v_{F}^{2}\right)^{2}+\lambda_{\Phi}(\Phi_{L}^{\dagger}\Phi_{L})^{2}
  +\lambda_{\Phi}(\Phi_{R}^{\dagger}\Phi_{R})^{2}+\lambda_{\phi_{c}}(\phi_{c}^{\dagger}\phi_{c})^{2}
  +\lambda_{\phi^{-}}(\phi^{+}\phi^{-})^{2}\nonumber\\
 &+2\lambda_{1}H_{L}^{\dagger}H_{L}H_{R}^{\dagger}H_{R}+2(H_{L}^{\dagger}H_{L}+H_{R}^{\dagger}H_{R})
  (\lambda_{2}\Phi_{L}^{\dagger}\Phi_{L}+\lambda_{2}\Phi_{R}^{\dagger}\Phi_{R}+\lambda_{3}\phi_{c}^{\dagger}\phi_{c}
  +\lambda_{4}\phi^{+}\phi^{-})\nonumber\\
 &+\lambda_{5}(\Phi_{L}^{\dagger}H_{L}H_{R}^{\dagger}\Phi_{R}+h.c.)
  +\lambda_{6}(\Phi_{L}^{\dagger}\widetilde{H}_{L}\widetilde{H}_{R}^{\dagger}\Phi_{R}+h.c.)\nonumber\\
 &+\mbox{other weak coupling terms},
\end{alignat}
 where $\widetilde{H}_{L}=-i\tau_{2}H_{L}^{*}$ and $\widetilde{H}_{R}^{\dagger}=H_{R}^{T}i\tau_{2}$. All of the self-coupling parameters, $\lambda_{H},\lambda_{F},\lambda_{\Phi},\lambda_{\phi_{c}},\lambda_{\phi^{-}}$, are positive and should be $\sim0.1$, while the interactive couplings, $\lambda_{1},\lambda_{2},\ldots,\lambda_{6}$, are weak and should be $<0.1$. $v_{L}$ and $v_{R}$ are respectively the vacuum expectation values (VEVs) of $H_{L}$ and $H_{R}$. The potential vacuum configurations are as follows,
\begin{alignat}{1}
 &\langle H_{L}\rangle =\frac{v_{L}}{\sqrt{2}}\left(\begin{array}{c}1\\0\end{array}\right),\hspace{0.5cm}
  \langle H_{R}\rangle =\frac{v_{R}}{\sqrt{2}}\left(\begin{array}{c}1\\0\end{array}\right),\nonumber\\
 &\langle F\rangle=v_{F}\mathrm{diag}
  \left(cos{\alpha}cos{\beta}e^{i\delta_{1}},cos{\alpha}sin{\beta}e^{i\delta_{2}},sin{\alpha}e^{i\delta_{3}}\right)
  =v_{F}Y_{F},
\end{alignat}
 where $\langle F\rangle$ is derived from the three complex scalar breakings. Explicitly, $v_{L}\neq v_{R}\neq 0$ will spontaneously break the gauge symmetries and the left-right mirror symmetry, on the other hand, $v_{F}\neq 0$ will spontaneously break the flavor symmetries, simultaneously, the $CP$ invariance will spontaneously be broken due to the non-vanishing and irremovable phases in $\langle F\rangle$. However, the symmetry breaking sequence is controlled by the hierarchy of the VEVs such as $v_{L}\sim10^{2}\ll v_{R}\sim10^{13}<v_{F}\sim10^{15}$ (in GeV as unit). Firstly, the flavor symmetries are broken and $CP$ is violated at the scale $v_{F}$. Secondly, at the scale $v_{R}$ $SU(2)_{R}\otimes U(1)_{B-L}$ is broken down $U(1)_{Y}$ which is namely the hypercharge subgroup of the SM, at the same time the left-right mirror symmetry is lost. Lastly, $SU(2)_{L}\otimes U(1)_{Y}\rightarrow U(1)_{em}$ is completed at the scale $v_{L}$, i.e. the electroweak breaking. These symmetry breakings are accomplished step by step in the universe evolution.

 The breaking of the flavor symmetries directly generates the Dirac mass of the Majorana fermions $N_{L}$ and $N_{R}$ as follows,
\ba
 M_{ND}=v_{F}Y_{5}^{\dagger}Y_{F}Y_{5}=v_{F}Y_{ND}\,.
\ea
 The Dirac mass term is the only source of the flavor breaking and $CP$ violation, and it will play a key role in origin of masses and mixings of the quarks and leptons.

 After the gauge symmetries are broken, the masses and mixings of the scalar and gauge bosons are easily derived by the standard program. The detailed expressions are as follows,
\begin{alignat}{1}
 &M_{H_{L}}=\sqrt{2\lambda_{H}}\,v_{L},\hspace{0.5cm} M_{H_{R}}=\sqrt{2\lambda_{H}}\,v_{R},\nonumber\\
 &M_{\Phi_{L,R}}=\sqrt{\lambda_{2}(v_{L}^{2}+v_{R}^{2})}\,,\hspace{0.3cm}
  M_{\phi_{c}}=\sqrt{\lambda_{3}(v_{L}^{2}+v_{R}^{2})}\,,\hspace{0.3cm} M_{\phi^{-}}=\sqrt{\lambda_{4}(v_{L}^{2}+v_{R}^{2})}\,,\nonumber\\
 &M_{W_{L}}=\frac{g_{w}v_{L}}{2}\,,\hspace{0.3cm} M_{Z_{L}}=\frac{M_{W_{L}}}{cos\theta_{W}}\,,\hspace{0.3cm}
  M_{W_{R}}=\frac{g_{w}v_{R}}{2}\,,\hspace{0.3cm} M_{Z_{R}}=\frac{M_{W_{R}}}{cos\theta_{W}'}\,,\nonumber\\
 &tan\theta_{W}'=\frac{g_{x}}{g_{w}}\,,\hspace{0.5cm} tan\theta_{W}=\frac{g_{Y}}{g_{w}}=sin\theta_{W}',\hspace{0.5cm}
  Q_{e}=I^{L}_{3}+I^{R}_{3}+\frac{B-L}{2}\,,
\end{alignat}
 where $\theta_{W}'$ and $\theta_{W}$ are the mixing angles of the neutral gauge fields for the $SU(2)_{R}$ breaking and the $SU(2)_{L}$ one, respectively. However, the mixings between the left-type bosons and the right-type ones are negligible since $v_{L}\ll v_{R}$. Obviously, the non-SM particles in (10) have super-heavy masses close to the scale $v_{R}$.

 On the basis of the model Lagrangian and symmetry breakings, the fermion masses and mixings are generated by the following mechanism. First of all, the super-heavy Majorana fermions $N_{L}$ and $N_{R}$ are decoupling below the scale $v_{F}$, so we can integrate them out and derive three types of the effective Yukawa couplings, i) the respective Majorana coupling of $l_{L}$ and $l_{R}$, ii) the Dirac coupling of $l_{L}$ and $l_{R}$, iii) two Dirac couplings of $q_{L}$ and $q_{R}$. The neutrino Dirac coupling is generated by (a) in Figure 1, and (b) in Figure 1 can generate the up-type quark coupling and the down-type one, which are respectively related to $H_{L},H_{R}$ and $\widetilde{H}_{L},\widetilde{H}_{R}$.
\begin{figure}
 \centering
 \includegraphics[totalheight=4.8cm]{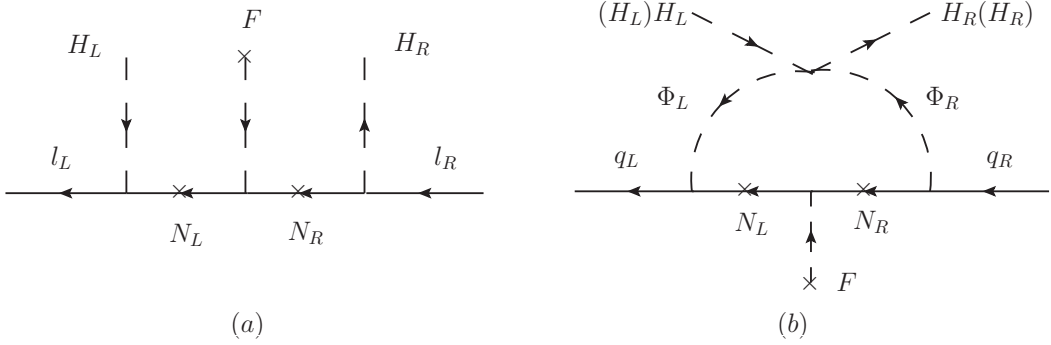}
 \caption{The generations of the fermion Dirac couplings, (a) for the neutrinos, (b) for the up-type and down-type quarks.}
\end{figure}
 These effective couplings will give the neutrino and quark masses after $H_{L}$ and $H_{R}$ developing the VEVs. Secondly, we can put together the neutrino Dirac coupling and the Yukawa couplings involving $\phi_{c}$ and $\phi^{-}$, thus a new  down-type quark coupling and a charged lepton coupling are generated by (a) in Figure 2. we can also make use of the up-type quark coupling and the Yukawa coupling involving $\phi_{c}$ to generate the other charged lepton coupling by (b) in Figure 2.
\begin{figure}
 \centering
 \includegraphics[totalheight=4cm]{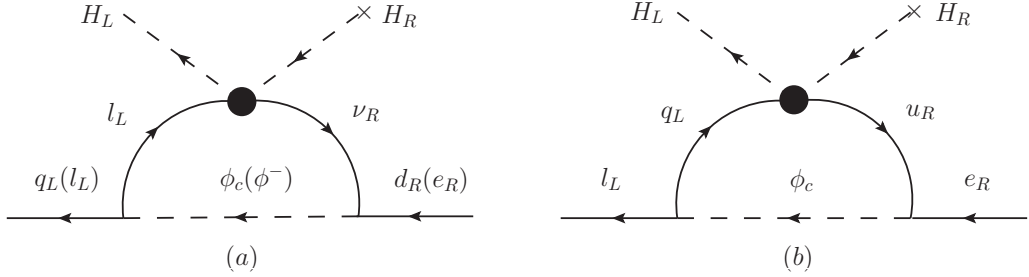}
 \caption{The generations of the fermion Dirac couplings, (a) for the down-type quarks and the charged leptons, (b) for the charged leptons.}
\end{figure}
 The black dots in Figure 2 namely denote the corresponding effective couplings in Figure 1. To sum up, the complete effective Yukawa couplings are given by
\begin{alignat}{1}
 \mathscr{L}^{eff}_{Yukawa}=
 &\:\frac{1}{2\overline{M}_{N}}\overline{l_{L}}H_{L}Y_{1}^{\dagger}Y_{N}^{*-1}Y_{1}^{*}H_{L}^{T}C\overline{l_{L}}^{T}
   +\frac{1}{2\overline{M}_{N}}l_{R}^{T}CH_{R}^{*}Y_{1}^{T}Y_{N}^{-1}Y_{1}H_{R}^{\dagger}l_{R}\nonumber\\
 &-\frac{v_{F}}{\overline{M}_{N}^{2}}\overline{l_{L}}H_{L}Y_{1}^{\dagger}Y_{N}^{*-1}Y_{ND}Y_{N}^{-1}Y_{1}H_{R}^{\dagger}l_{R}
  +\frac{\lambda_{5}B_{0}v_{F}}{16\pi^{2}\overline{M}_{N}^{2}}\overline{q_{L}}H_{L}Y_{2}^{\dagger}
   Y_{N}^{*-1}Y_{ND}Y_{N}^{-1}Y_{2}H_{R}^{\dagger}q_{R}\nonumber\\
 &+\frac{\lambda_{6}B_{0}v_{F}}{16\pi^{2}\overline{M}_{N}^{2}}\overline{q_{L}}\widetilde{H}_{L}Y_{2}^{\dagger}
   Y_{N}^{*-1}Y_{ND}Y_{N}^{-1}Y_{2}\widetilde{H}_{R}^{\dagger}q_{R}
  +\frac{\sqrt{2}}{16\pi^{2}v_{R}}\overline{q_{L}}\widetilde{H}_{L}Y_{3}^{\dagger}
  (A_{\alpha\beta}Y_{D\alpha\beta}^{*})Y_{3}\widetilde{H}_{R}^{\dagger}q_{R}\nonumber\\
 &+\frac{\sqrt{2}}{16\pi^{2}v_{R}}\overline{l_{L}}\widetilde{H}_{L}Y_{4}^{\dagger}
  (B_{\alpha\beta}Y_{D\alpha\beta}^{*})Y_{4}\widetilde{H}_{R}^{\dagger}l_{R}
  +\frac{\sqrt{2}}{16\pi^{2}v_{R}}\overline{l_{L}}\widetilde{H}_{L}Y_{3}^{*}
  (C_{\alpha\beta}Y_{u\alpha\beta}^{*})Y_{3}^{T}\widetilde{H}_{R}^{\dagger}l_{R}\,,
\end{alignat}
 where $Y_{D}$ and $Y_{u}$ are respectively the effective couplings of the neutrinos and the up-type quarks, which are directly derived from the second line of (11), see the following (13). In (11), all kinds of the loop integration factors are eventually simplified by the two-point functions as follows,
\begin{alignat}{1}
 &B_{0}=B_{0}(M_{H_{L}}^{2},M_{\Phi_{L}}^{2},M_{\Phi_{R}}^{2}),\nonumber\\
 &A_{\alpha\beta}=B_{0}(M_{H_{L}}^{2},m_{\nu_{L\alpha}}^{2},M_{\nu_{R\beta}}^{2})
  -B_{0}(m_{d}^{2},M_{\nu_{R\beta}}^{2},M_{\phi_{c}}^{2})-B_{0}(m_{d}^{2},m_{\nu_{L\alpha}}^{2},M_{\phi_{c}}^{2}),\nonumber\\
 &B_{\alpha\beta}=B_{0}(M_{H_{L}}^{2},m_{\nu_{L\alpha}}^{2},M_{\nu_{R\beta}}^{2})
  -B_{0}(m_{e}^{2},M_{\nu_{R\beta}}^{2},M_{\phi^{-}}^{2})-B_{0}(m_{e}^{2},m_{\nu_{L\alpha}}^{2},M_{\phi^{-}}^{2}),\nonumber\\
 &C_{\alpha\beta}=B_{0}(M_{H_{L}}^{2},m_{u_{\alpha}}^{2},m_{u_{\beta}}^{2})
  -B_{0}(m_{e}^{2},m_{u_{\beta}}^{2},M_{\phi_{c}}^{2})-B_{0}(m_{e}^{2},m_{u_{\alpha}}^{2},M_{\phi_{c}}^{2}),
\end{alignat}
 where $\alpha,\beta$ are the generation indexes, they are not summed in (11).

 After the gauge symmetry breakings, (11) finally leads to the lepton and quark masses as follows,
\begin{alignat}{1}
 &M_{L}=-\frac{v_{L}^{2}}{2\overline{M}_{N}}Y_{1}^{\dagger}Y_{N}^{*-1}Y_{1}^{*},\hspace{0.5cm}
  M_{R}=-\frac{v_{R}^{2}}{2\overline{M}_{N}}Y_{1}^{T}Y_{N}^{-1}Y_{1},\nonumber\\
 &M_{D}=\frac{v_{F}v_{L}v_{R}}{2\overline{M}_{N}^{2}}Y_{1}^{\dagger}Y_{N}^{*-1}Y_{ND}Y_{N}^{-1}Y_{1},\hspace{0.3cm}
  M_{u}=-\frac{\lambda_{5}B_{0}v_{F}v_{L}v_{R}}{32\pi^{2}\overline{M}_{N}^{2}}Y_{2}^{\dagger}Y_{N}^{*-1}Y_{ND}Y_{N}^{-1}Y_{2},\nonumber\\
 &M_{d}=\frac{\lambda_{6}}{\lambda_{5}}M_{u}+\frac{1}{16\pi^{2}}Y_{3}^{\dagger}(A_{\alpha\beta}M^{*}_{D\alpha\beta})Y_{3},\nonumber\\
 &M_{e}=\frac{1}{16\pi^{2}}Y_{4}^{\dagger}(B_{\alpha\beta}M^{*}_{D\alpha\beta})Y_{4}
  +\frac{1}{16\pi^{2}}Y_{3}^{*}(C_{\alpha\beta}M^{*}_{u\alpha\beta})Y_{3}^{T},\nonumber\\
 &M_{\nu_{L}}^{eff}=M_{L}-M_{D}M_{R}^{-1}M_{D}^{T},\nonumber\\
 &Y_{f=D,u,d,e}=-\frac{\sqrt{2}}{v_{L}}M_{f},
\end{alignat}
 where the effective mass of the left-handed neutrinos is implemented by the see-saw mechanism \cite{13}, in addition, the effective Yukawa couplings at the electroweak scale are related to the fermion masses by the last equation in (13). This set of equations of (13) clearly shows the origin of the fermion masses and the interrelations among them. The neutrino Majorana masses are derived from the coupling $Y_{N}$. If the flavor symmetries are unbroken, the Dirac coupling $Y_{ND}$ is vanishing, then the Dirac masses $M_{f=D,u,d,e}$ are all nought. Therefore, $Y_{N}$ and $Y_{ND}$ are respectively the roots of the Majorana and Dirac masses, all kinds of the fermion masses stem from them, in particular, $Y_{ND}$ is the only source of the flavor symmetry breaking and the $CP$ violation.

 For fitting the experimental data, the system of equations (13) are very difficult to be solved. As has been noted early, $M_{L},M_{R},M_{D},M_{u}$ are all symmetric matrices, in addition, $M_{L},M_{R}$ can be diagonalized by the following unitary matrix $U_{0}$ owing to the flavor symmetries but two of the eigenvalues are degenerate. However, we can choose the flavor basis such that $M_{D}$ and $M_{u}$ are diagonal matrices simultaneously, then solving becomes relatively easy. Lastly, $Y_{4}$ is relatively smaller so as to implement the following leptogenesis, so the second term in $M_{e}$ actually makes major contributions. In this simple scenario, the fermion masses are newly expressed as follows,
\begin{alignat}{1}
 &M_{L}=\frac{v_{L}^{2}}{v_{R}^{2}}M_{R}^{*},\hspace{0.3cm}
  M_{R}=U_{0}\mathrm{diag}(M_{\nu_{R1}}=M_{\nu_{R2}},M_{\nu_{R3}})U_{0}^{T},\hspace{0.3cm}
  U_{0}=\frac{1}{\sqrt{6}}\left(\begin{array}{ccc}2&0&\sqrt{2}\\-1&\sqrt{3}&\sqrt{2}\\-1&-\sqrt{3}&\sqrt{2}\end{array}\right),\nonumber\\
 &M_{D}=U_{D}\mathrm{diag}(m_{D1},m_{D2},m_{D3})U_{D}^{T},\hspace{0.5cm}
  M_{u}=U_{u}\mathrm{diag}(m_{u},m_{c},m_{t})U_{u}^{T},\nonumber\\
 &M_{d}'=\frac{\lambda_{6}}{\lambda_{5}}M_{u}'+\frac{1}{16\pi^{2}}Y_{3}'^{T}\mathrm{diag}(A_{11},A_{22},A_{33})M_{D}'Y_{3}'
  =U_{CKM}\mathrm{diag}(m_{d},m_{s},m_{b})U_{CKM}^{T},\nonumber\\
 &M_{e}'\approx\frac{1}{16\pi^{2}}Y_{3}'\mathrm{diag}(C_{11},C_{22},C_{33})M_{u}'Y_{3}'^{T}
  =U_{e}\mathrm{diag}(m_{e},m_{\mu},m_{\tau})U_{e}^{T},\nonumber\\
 &M_{\nu_{L}}'=U_{D}^{\dagger}M_{L}U_{D}^{*}-M_{D}'U_{D}^{T}M_{R}^{-1}U_{D}M_{D}'^{T}
  =U_{e}U_{MNS}\mathrm{diag}\left(m_{\nu_{L1}},m_{\nu_{L2}},m_{\nu_{L3}}\right)U_{MNS}^{T}U_{e}^{T},\nonumber\\
 &Y_{3}'=U_{D}^{\dagger}Y_{3}U_{u}^{*}=4\pi U_{e}
  \mathrm{diag}(\sqrt{\frac{m_{e}}{C_{11}m_{u}}},\sqrt{\frac{m_{\mu}}{C_{22}m_{c}}},\sqrt{\frac{m_{\tau}}{C_{33}m_{t}}}),
\end{alignat}
 where the superscript apostrophe means that the masses and couplings are in the new flavor basis. $U_{CKM}$ and $U_{MNS}$ are respectively the quark mixing matrix and the lepton one, which were defined in \cite{14} and \cite{15}. For all kinds of the unitary matrices, their mixing angles and $CP$-violating phases are parameterized by the standard form in particle data group \cite{16}. The solution of $Y_{3}'$ has directly been derived from the equation of $M_{e}'$. If the coupling with $Y_{3}$ is vanishing, then the quark mixing $U_{CKM}$ will become an unit matrix, and the interrelations between the quark masses and the lepton ones are nothing. In (14), most of the mass and mixing parameters are fixed by the experimental data, the undetermined parameters include five mass parameters $M_{\nu_{R2}},M_{\nu_{R3}},m_{D1},m_{D2},m_{D3}$, two mixing matrices $U_{e},U_{D}$, and three scalar sector parameters $v_{R},\frac{\lambda_{5}}{\lambda_{6}},M_{\phi_{c}}$, where $M_{\phi_{c}}$ is involved in $A_{\alpha\beta}$ and $C_{\alpha\beta}$. Now the difficulty of solving (14) is greatly reduced. In conclusion, all of the above contents are the theoretical framework of the model.

\vspace{1cm}
 \noindent\textbf{III. Leptogenesis}

\vspace{0.3cm}
 After the flavor symmetries are broken and the $CP$ is violated below the scale $v_{F}$, the universe undergoes inflation and reheating. The reheating temperature can reach to $T_{reheat}\sim 10^{13-14}$ GeV in some of the inflation models \cite{17}. At this scale $v_{R}\sim T_{reheat}$ the $SU(2)_{R}\otimes U(1)_{B-L}$ gauge symmetry is broken and the left-right mirror symmetry is lost. As a result, the non-SM gauge and scalar bosons and the right-handed neutrinos $\nu_{R}$ obtain their masses. In virtue of the lepton number violation, $\nu_{R}$ become the Majorana neutrinos. As the universe expansion and cooling, the charged scalar $\phi^{-}$ can decay into $l_{L}+l_{L}$ or $e_{R}+\nu_{R}$. However, the decay $\phi^{-}\rightarrow e_{R}+\nu_{R}$ evidently violates two units of the lepton number. In view of the $CP$ violation in the effective Yukawa couplings, a $CP$ asymmetry of the decay process is generated through the interference between the tree diagram and the loop one, shown as figure 3.
\begin{figure}
 \centering
 \includegraphics[totalheight=4.5cm]{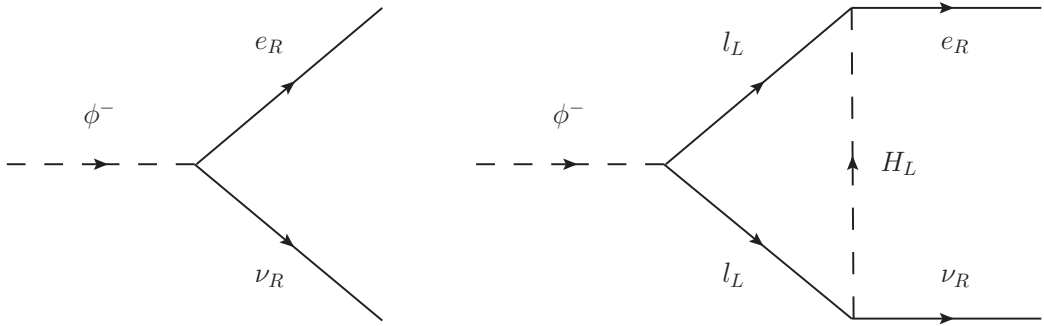}
 \caption{The graphs of the decay $\phi^{-}\rightarrow e_{R}+\nu_{R}$ which leads to leptogenesis.}
\end{figure}
 In addition, the $\phi^{-}$ decay is actually an out-of-equilibrium process for an enough small $Y_{4}$, namely the decay rate is far smaller than the Hubble expansion rate of the universe,
\ba
 \Gamma(\phi^{-}\rightarrow e_{R}+\nu_{Ri})=\frac{M_{\phi^{-}}(Y_{4}Y_{4}^{\dagger})_{ii}}{16\pi}
 \ll H(T=M_{\phi^{-}})=\frac{1.66\sqrt{g_{*}}M_{\phi^{-}}^{2}}{M_{pl}}\,,
\ea
 where $M_{pl}=1.22\times10^{19}$ GeV, and $g_{*}$ is an effective number of relativistic degrees of freedom at $T=M_{\phi^{-}}$. At this temperature, the relativistic particles are exactly the SM ones, so one can easily figure out $g_{*}=106.75$. In short, the decay $\phi^{-}\rightarrow e_{R}+\nu_{R}$ can satisfy three Sakharov's conditions \cite{18}, so the baryogenesis can be implemented through the leptogenesis in the model.

 The decay $CP$ asymmetry is defined and calculated as follow,
\begin{alignat}{1}
 &\varepsilon=\frac{\sum\limits_{i}[\Gamma(\phi^{-}\rightarrow e_{R}+\nu_{Ri})
  -\Gamma(\phi^{+}\rightarrow \overline{e_{R}}+\overline{\nu_{Ri}})]}{\Gamma_{total}(\phi^{-})}
 =\frac{Im[\sum\limits_{i}(Y_{D}^{T}Y_{4}Y_{e}Y_{4}^{\dagger})_{ii}f_{i}]}{4\pi(Tr[Y_{4}Y_{4}^{\dagger}]+\sum\limits_{i}(Y_{4}Y_{4}^{\dagger})_{ii})}\,,\nonumber\\
 &\Gamma_{total}(\phi^{-})=\Gamma(\phi^{-}\rightarrow l_{L}+l_{L})
  +\sum_{i}\Gamma(\phi^{-}\rightarrow e_{R}+\nu_{Ri}),\hspace{0.5cm}
  f_{i}=1-2\frac{M^{2}_{\nu_{Ri}}}{M^{2}_{\phi^{-}}}\,,
\end{alignat}
 where the sum index $i$ is limited by $M_{\nu_{Ri}}<M_{\phi^{^{-}}}$. (16) clearly manifests that $\varepsilon$ is closely related to the lepton $CP$ violation, which is directly indicated by $Y_{D}$ and $Y_{e}$. The size of $\varepsilon$ is essentially dominated by $Y_{D}$ and $Y_{e}$, while $Y_{4}$ has only little influence on it. However, (16) can naturally give a required value of $\varepsilon$, namely $\varepsilon\sim10^{-8}$.

 Below the scale $v_{R}$, the baryon number is conserved but the lepton number is violated, the above-mentioned decay process accordingly leads to the $B-L$ asymmetry in the universe. Afterwards, the $B-L$ asymmetry is partly converted into the baryon asymmetry through the sphaleron processes above the electroweak scale \cite{19}. These results are expressed as follow,
\ba
 Y_{B-L}=\frac{n_{B-L}-\overline{n}_{B-L}}{s}=\kappa\frac{(-2)\varepsilon}{g_{*}}\,,\hspace{0.3cm}
 Y_{B}=c_{s}Y_{B-L},\hspace{0.3cm} \eta_{B}=\frac{n_{B}-\overline{n}_{B}}{n_{\gamma}}=7.04Y_{B}.
\ea
 $s$ is the entropy density. $\kappa$ is a dilution factor, it can be approximated to $\kappa\approx1$ for a very weak decay which is serious departure from thermal equilibrium. $c_{s}=\frac{28}{79}$ is a coefficient of the electroweak sphaleron conversion. $7.04$ is a ratio of the entropy density $s$ to the photon number density $n_{\gamma}$. Obviously, the value of $\eta_{B}$ is completely determined by $\varepsilon$. Therefore, the present-day baryon asymmetry can be explained in the model very well.

\vspace{1cm}
 \noindent\textbf{IV. Numerical Results}

\vspace{0.3cm}
 In the section I present the model numerical results. The model involves a lot of the new parameters besides the SM ones. In principle the SM parameters are fixed by the experimental data, the non-SM parameters will be determined by solving the system of equation (14) and fitting the baryon asymmetry. In the light of the foregoing discussions, we can choose the below quantities of the SM as the input parameters (mass unit in GeV),
 \begin{alignat}{1}
 &sin\theta_{W}=0.231,\hspace{0.5cm} M_{Z_{L}}=91.2,\hspace{0.5cm}M_{H_{L}}=126,\hspace{0.5cm} v_{L}=246,\nonumber\\
 &m_{u}=0.0023,\hspace{0.5cm} m_{c}=1.275,\hspace{0.5cm} m_{t}=173,\nonumber\\
 &m_{e}=0.000511,\hspace{0.5cm} m_{\mu}=0.1057,\hspace{0.5cm} m_{\tau}=1.777.
\end{alignat}
 All of them come from the particle data group \cite{16}. The non-SM parameters are input by the below values,
 \begin{alignat}{1}
 &v_{R}=2\times10^{14}\:\mathrm{GeV},\hspace{0.5cm} M_{\phi_{c}}=2\times10^{13}\:\mathrm{GeV},\hspace{0.5cm} \frac{\lambda_{5}}{\lambda_{6}}=0.0242,\nonumber\\
 &M_{\nu_{R1}}=M_{\nu_{R2}}=6.05\times10^{12}\:\mathrm{GeV},\hspace{0.5cm}M_{\nu_{R3}}=3.55\times10^{13}\:\mathrm{GeV},\nonumber\\
 &(Y_{D1},Y_{D2},Y_{D3})=0.0645\times(0,-1,1),\nonumber\\
 &sin\theta_{12}^{e}=0.058,\hspace{0.5cm} sin\theta_{23}^{e}=0.66,\hspace{0.5cm} sin\theta_{13}^{e}=0.0313,\hspace{0.5cm}
  \delta^{e}=-0.48\pi,\nonumber\\
 &sin\theta_{12}^{D}=\frac{1}{2},\hspace{0.5cm} sin\theta_{23}^{D}=0.805,\hspace{0.5cm}
  sin\theta_{13}^{D}=\frac{1}{\sqrt{2}},\hspace{0.5cm} \delta^{D}=0.11\pi,
\end{alignat}
 where the last two lines are respectively the mixing angles and $CP$-violating phases in $U_{e}$ and $U_{D}$. In (19), $v_{R}$ and $M_{\phi_{c}}$ are actually taken as the fixed values. Firstly, the parameters, $\frac{\lambda_{5}}{\lambda_{6}},Y_{D1},Y_{D2},Y_{D3},U_{e}$, are determined by the equation of $M_{d}'$ in (14). Secondly, the remaining parameters, $M_{\nu_{R1}},M_{\nu_{R3}},U_{D}$, are solved by the equation of $M_{\nu_{L}}'$ in (14). In this fitting, it is however found that $(Y_{D1},Y_{D2},Y_{D3})$ can be set as only one tuning parameter and $sin\theta_{12}^{D}=\frac{1}{2}, sin\theta_{13}^{D}=\frac{1}{\sqrt{2}}$ can also be fixed. Therefore, the adjustable parameters in (19) are in fact ten in all. Of course, the parameter characteristics arise from the model theoretical structures.

 Firstly, by use of (10) we can obtain the masses of the right-type gauge and scalar bosons (in GeV as unit),
\ba
 M_{W_{R}}=6.5\times10^{13},\hspace{0.5cm} M_{Z_{R}}=7.8\times10^{13},\hspace{0.5cm} M_{H_{R}}=1.02\times10^{14}.
\ea
 They are mainly subject to $v_{R}$. However, these particles are too heavy and irrelevant to the present low-energy phenomenology.

 Secondly, by use of (14) we can predict the down-type quark masses, the quark mixing $U_{CKM}$, the light neutrino masses, and the lepton mixing $U_{MNS}$. All kinds of results are given as follows,
\begin{alignat}{1}
 &m_{d}=4.9\:\mathrm{MeV},\hspace{0.5cm} m_{s}=95\:\mathrm{MeV},\hspace{0.5cm} m_{b}=4.18\:\mathrm{GeV},\nonumber\\
 &sin\theta_{12}^{ckm}=0.2252,\hspace{0.5cm} sin\theta_{23}^{ckm}=0.0415,\hspace{0.5cm}
  sin\theta_{13}^{ckm}=0.00352,\nonumber\\
 &\delta^{ckm}=-0.397\pi,\hspace{0.5cm} J_{cp}^{ckm}=-3.03\times10^{-5},\nonumber\\
 &m_{\nu_{L1}}=0.0079\:\mathrm{eV},\hspace{0.5cm} m_{\nu_{L2}}=0.012\:\mathrm{eV},\hspace{0.5cm}
  m_{\nu_{L3}}=0.05\:\mathrm{eV},\nonumber\\
 &\triangle m^{2}_{21}=7.5\times10^{-5}\:\mathrm{eV^{2}},\hspace{0.5cm}
  \triangle m^{2}_{32}=2.4\times10^{-3}\:\mathrm{eV^{2}},\nonumber\\
 &sin\theta_{12}^{mns}=0.552,\hspace{0.5cm} sin\theta_{23}^{mns}=0.707,\hspace{0.5cm}
  sin\theta_{13}^{mns}=0.156,\nonumber\\
 &\delta^{mns}=0.975\pi,\hspace{0.5cm} J_{cp}^{mns}=2.75\times10^{-3},\hspace{0.5cm}
  \beta^{mns}_{1}=-0.53\pi,\hspace{0.5cm} \beta^{mns}_{2}=-0.5\pi,
\end{alignat}
 where $J_{cp}^{ckm}$ and $J_{cp}^{mns}$ are respectively the $CP$-violating Jarlskog invariants in the quark and lepton sectors, $\beta_{1}^{mns}$ and $\beta_{2}^{mns}$ are two Majorana phases in $U_{MNS}$. Obviously, all the results are very well in accordance with the current experimental data \cite{16}. In particular, we can use the fewer parameters to predict the greater mass and mixing values, thus the model shows a large power of predictions. This success is of course owing to the correlations of the fermion mass matrices, which originate from the model symmetries and their breakings. (21) predicts that the lepton $CP$ violation is relatively larger, so it will possibly be detected in near future.

 Lastly, we also need input $M_{\phi^{-}}$ and $Y_{4}$ in order to calculate the baryon asymmetry. Their values are taken as
\ba
 M_{\phi^{-}}=9.3\times10^{12}\:\mathrm{GeV},\hspace{0.5cm} Y_{4}=10^{-3}(T-T^{T}),\hspace{0.5cm} Y_{4}'=U_{D}^{\dagger}Y_{4}U_{D}^{*},
\ea
 where $Y_{4}$ is fixed and only $M_{\phi^{-}}$ is tuned to fit $\eta_{B}$. It should be noted that $M_{\nu_{R1}}=M_{\nu_{R2}}<M_{\phi^{-}}<M_{\nu_{R3}}$, so the sum in (16) is only for $i=1,2$. By use of (15)-(17), the out-of-equilibrium condition and the baryon asymmetry are calculated as
\ba
 \frac{\Gamma(\phi^{-}\rightarrow e_{R}+\nu_{R1,2})}{H(T=M_{\phi^{-}})}=0.0088,\hspace{0.5cm}
 \eta_{B}=6.12\times10^{-10}.
\ea
 The above first relation clearly demonstrates that the $\phi^{-}$ decay is indeed out-of-equilibrium. The value of $\eta_{B}$ is precisely in agreement with the current data of the universe observations \cite{20}. In brief, all of the numerical results are naturally produced without any fine tuning. They not only correctly reproduce all kinds of the experimental data, but also give some predictions. This clearly demonstrates that the model is reasonable and feasible.

\vspace{1cm}
 \noindent\textbf{V. Conclusions}

\vspace{0.3cm}
 In this paper, I discuss the left-right symmetric model with the gauge groups $SU(3)_{C}\otimes SU(2)_{L}\otimes SU(2)_{R}\otimes U(1)_{B-L}$ and the flavor symmetries $Z_{3L}\otimes Z_{3R}$. The model can account for the origin of the fermion masses and $CP$ violation, and also implement the baryogenesis through the leptogenesis. At first, the flavor symmetries and the $CP$ invariance are spontaneously broken by the flavor field developing the VEV, soon afterwards, the right-handed isospin and $B-L$ gauge subgroups are spontaneously breakings, and the left-right mirror symmetry is lost. It is these symmetries and their breakings that lead to the special effective Yukawa couplings at the low energy, which eventually generate the fermion masses and mixings. In the light of the model theoretical structures, the lepton and quark masses have common origin, namely all of them stem from the Majorana and Dirac masses of $N_{L}$ and $N_{R}$, in particular, the Dirac coupling $Y_{ND}$ is the only source of the flavor breaking and $CP$ violation. As a result, the fermion mass matrices are not independent but rather there are the correlations among them. The model can naturally and correctly reproduce all kinds of masses and mixings of the SM and neutrino physics by the fewer parameters, so it shows a large power of predictions. In addition, the decay of the super-heavy scalar $\phi^{-}\rightarrow e_{R}+\nu_{R}$ satisfies the lepton number violation, the $CP$ asymmetry, and being out-of-equilibrium. The $CP$ asymmetry is directly related to the lepton $CP$ violation. The generated lepton asymmetry is eventually converted into the baryon asymmetry through the sphaleron processes above the electroweak scale. All of the numerical results are in agreement with the experimental data very well. The model predicts the leptonic $J_{cp}^{mns}\approx2.75\times10^{-3}$, it will possibly be detected in near future. Finally, the model is feasible and promising to be tested in future experiments.

\vspace{1cm}
 \noindent\textbf{Acknowledgments}

\vspace{0.3cm}
 I would like to thank my wife for her large helps. This research is supported by the Fundamental Research Funds for the Central Universities Grant No. WK2030040054.

\end{document}